\documentclass[11pt,twoside]{article}

\usepackage{asp2006}
\usepackage{epsf}
\usepackage{psfig}
\usepackage{lscape}
\usepackage{graphicx}
\usepackage{natbib}

\begin{document}
\title{Magnetic topologies of the Herbig Ae/Be stars}   %%% Fill in title
\author{E.~Alecian,$^1,2$ G.A.~Wade,$^1$ C.~Catala, $^2$, S.~Bagnuo,$^3$ T.~B\"ohm,$^4$ J.-C.~Bouret,$^5$ J.-F.~Donati,$^4$ C.~Folsom,$^3$ J.~Grunhut,$^1$ J.D.~Landstreet,$^6$ P.~Petit,$^4$ J.~Silvester$^1$}   %%% Fill in author names
\affil{$^1$Department of Physics, Royal Military College of Canada, PO Box 17000, Stn Forces, Kingston K7K 7B4, Canada}
\affil{$^2$Observatoire de Paris, LESIA, 5 place Jules Janssen, F-92195 Meudon Principal Cedex, France}
\affil{$^3$Armagh Observatory, College Hill, Armagh BT61 9DG}
\affil{$^4$Laboratoire dÕAstrophysique, Observatoire Midi-Pyr\'en\'ees, 14 avenue Edouard Belin, F-31400 Toulouse, France}
\affil{$^5$Laboratoire dÕAstrophysique de Marseille, Traverse du Siphon, BP8-13376 Marseille Cedex 12, France}
\affil{$^6$Department of Physics \& Astronomy, University of Western Ontario, London N6A 3K7, Canada}
    %%% Fill in author affiliations

\begin{abstract} %%% Abstract to run on from here.
Our recent discoveries of magnetic fields in a small number of Herbig Ae/Be (HAeBe) stars, the evolutionary progenitors of main sequence A/B stars, raise new questions about the origin of magnetic fields in the intermediate mass stars. The favoured fossil field hypothesis suggests that a few percent of magnetic pre-main sequence A/B stars should exhibit similar magnetic strengths and topologies to the magnetic Ap/Bp stars. In this talk I will present the methods that we have used to characterise the magnetic fields of the Herbig Ae/Be stars, as well as our first conclusions on the origin of magnetism in intermediate-mass stars. 
\end{abstract}

%%% MAIN BODY OF TEXT GOES HERE. CONSULT "INSTRUCTIONS FOR AUTHORS USING
%%% LATEX2E MARKUP", SECTIONS 2.3-2.6 FOR HELP WITH EQUATIONS, FIGURES,
%%% AND TABLES.

%\section{}   %%% Top level section head (remove "%" symbol)
%\subsection{}   %%% Second level section head (remove "%" symbol)
%\subsubsection{}   %%% Lowest level section head (remove "%" symbol)
%\section*{}    %%% Unnumbered top level section head (remove "%" symbol)
%\subsection*{}   %%% Unnumbered second level section head (remove "%" symbol)

\section{Introduction}

Around 5\% of the main sequence (MS) A and B stars host strong, large-scale magnetic fields. Such magnetic stars also show important abundance inhomogeneities in their photosphere, and are called the chemically peculiar Ap/Bp stars. The strength, topology and relationship to other physical characteristics of the magnetic fields of these stars cannot be explained by an envelope dynamo as in the sun. The most reliable hypothesis is the fossil field theory. This theory proposes that the observed magnetic fields of the MS Ap/Bp stars are relics of magnetic fields, either present in their parental molecular clouds, or generated during the very early stellar formation stages. This could imply that the magnetic fields survive through the whole life of these intermediate mass stars, from the pre-main sequence (PMS) phase to the white dwarf and neutron stars stages.

Until recently, magnetic fields were only directly observed in molecular clouds, MS A/B stars, and also in white dwarfs and neutron stars. In order to confirm the fossil field theory we need to observe fields at other intermediate stellar evolution stages, in particular during the PMS phase. With this aim we performed a large survey of ~130 intermediate mass PMS stars, the Herbig Ae/Be (HAeBe) stars, present in the field of the Galaxy and also in very young open clusters, using the new high-resolution spectropolarimeter ESPaDOnS. We attempted to detect magnetic fields in each star of the sample using the polarised properties of the Zeeman effect in their photospheres. We discovered 7 new magnetic HAeBe stars, leading to a calculated incidence of $\sim 6$~\% magnetic HAeBe stars (Wade et al. 2005, Catala et al. 2007, Alecian et al. 2008a, Alecian et al. 2008b, Wade et al. this proceeding). Taking into account the fossil field hypothesis and the incidence of the MS magnetic A/B stars (Power et al., in preparation), we predict between 1 and 10~\% of HAeBe stars could be magnetic. This result, while in agreement with the fossil field hypothesis, is not sufficient to totally confirm this theory. We need to go further.

According to this theory, magnetic fields among PMS stars must have similar topologies to the Ap/Bp stars, and they must show magnetic strengths consistent with magnetic flux conservation. We therefore need to characterise the magnetic fields of the HAeBe stars. In Section 2, I explain briefly our observations. In Section 3, I describe the method that we used to characterise their magnetic fields. The results and a discussion are presented in Section 4.

\section{Observations}

We obtained high-resolution Stokes $I$ and $V$ spectra of HAeBe stars using the spectropolarimeter ESPaDOnS (Donati et al. in prep.) installed at the 3.6m Canada-France-Hawaii telescope (CFHT), and using its twin, Narval, installed at the 2m Bernard Lyot Telescope (TBL) at the Pic du Midi Observatory (France). The data has been reduced using the Libre Esprit package especially developed for ESPaDOnS and Narval (Donati et al. 1997). For more information about the data see Wade et al. (this proceeding) and Alecian et al. (2008a).

\section{Characterisation of magnetic fields of Herbig Ae/Be stars}

In the following I will describe the general characterisation method that we applied for all magnetic HAeBe stars, by presenting the analysis that we performed on one of the magnetic HAeBe stars HD 200775.

\subsection{Extraction of LSD profiles}

During the PMS phase, stars are still contracting toward the MS. The stellar radius decreases by up to a factor of $\sim 4$. If magnetic flux is conserved, as predicted the fossil field theory, the magnetic strengths in their photospheres can be lower than in the Ap/Bp stars by around one order of magnitude. Furthermore the optical spectra of HAeBe stars are very often strongly contaminated by emission. For these reasons the detection of magnetic fields in the individual lines of HAeBe spectra is particularly challenging. We therefore need to apply the least squares deconvolution (LSD) method developed by Donati et al. (1997), that will allow us to increase the S/N ratio of our data by a factor up to 10.

\begin{figure}
\begin{center}
\psfig{figure=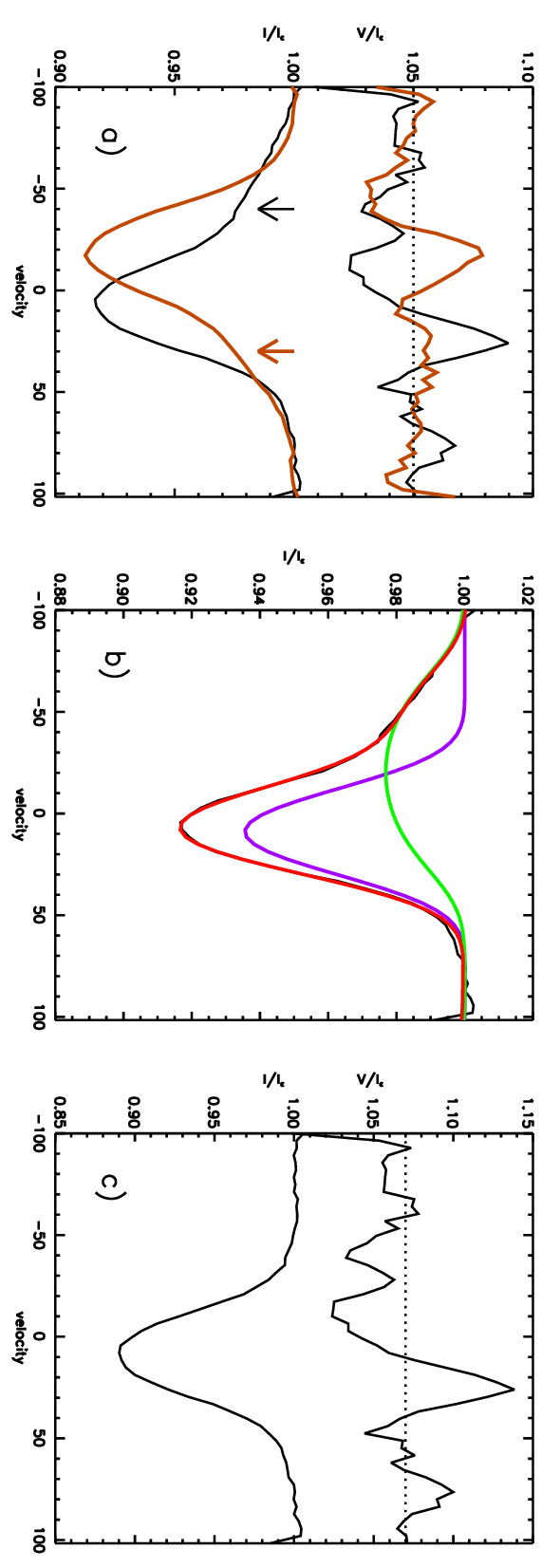,angle=90,width=13.5cm}
\caption{{\bf a)} LSD Stokes $I$ (lower) and $V$ (upper) profiles of HD 200775 (B3 V) observed in Sept. 2004 (black) and in June 2006 (red). The arrows indicate the radial velocity change of the secondary between both observations. {\bf b)}~Best model of the HD 200775 components (red) superimposed on the observed profile of Sept. 2004 (black). The individual synthetic profiles of the primary (blue) and secondary green) are also plotted. {\bf c)} The derived observed Stokes $I$ and $V$ profiles of the primary.}
\label{fig:lsd}
\end{center}
\end{figure}

The LSD technique increases considerably the S/N ratio of the data by combining the information contained in all the lines of the Stokes $I$ and $V$ spectra. This method assumes that all selected lines of the spectrum have a profile of similar shape. Hence, this supposes that all lines are broadened in the same way. We can therefore consider that the observed spectrum is a convolution between a profile (which is the same for all lines) and a mask encoding the positions and strengths of all chosen lines of the spectrum. We therefore apply a deconvolution to the observed spectrum using the mask, in order to obtain the average photospheric profiles of Stokes $I$ and $V$. In this procedure, each line is weighted by its S/N ratio, its depth in the unbroadened model and its Land\'e factor. The masks are computed using Kurucz ATLAS9 models (Kurucz 1993) with effective temperature and gravity suitable for each star. We excluded from this mask hydrogen Balmer lines, strong resonance lines and lines whose Land\'e factor is unknown. Exemple resulting LSD profiles for HD 200775 are shown in Fig. \ref{fig:lsd}a.

Figure \ref{fig:lsd}a shows a single Zeeman signature in the LSD Stokes $V$ profiles (upper profile), characteristic of a stellar magnetic field. The Stokes $I$ profile shows the profile of a binary star whose primary rotates with a $v\sin i=26 \pm2$~km/s and the secondary with $v\sin i=59 \pm 5$~km/s. A detail analysis of the radial velocity variations of the Stokes $I$ and $V$ profiles (illustrated in Fig. \ref{fig:lsd}a) leads to the conclusion that the Stokes $V$ signature comes from only the primary.

\begin{figure}
\begin{center}
\psfig{figure=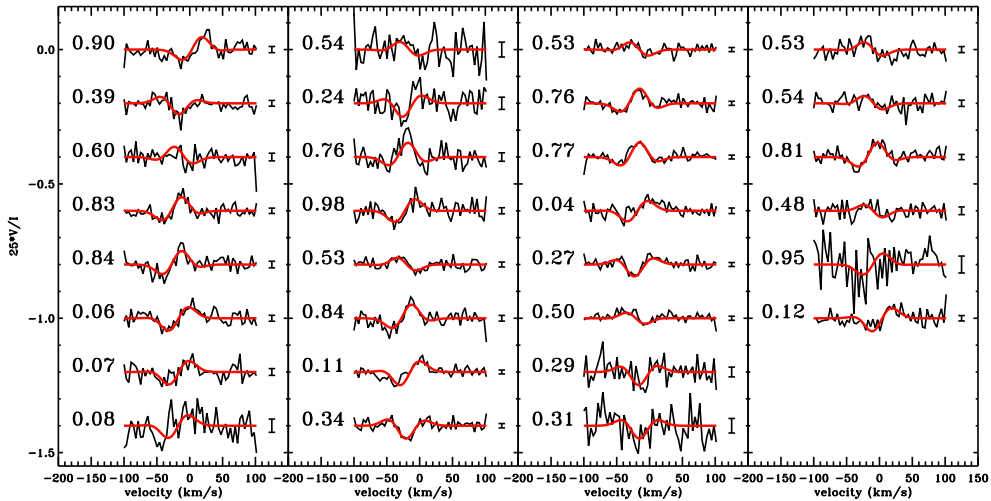,angle=90,width=13cm}
\caption{Stokes V profiles of HD 200775 (noisy black lines) superimposed to the synthetic ones (in red smooth lines ), corresponding to our best fit. The rotation phase and the error bars are indicated on the left and on the right of each profile, respectively (Alecian et al. 2008a).}
\label{fig:fitv}
\end{center}
\end{figure}

In order to determine the magnetic topology and strength of the primary we need to isolate the primary Stokes $I$ profile. In that aim we have fitted the binary profile using the sum of two synthetic single profiles. Each one of these profiles are computed using the convolution of a rotation profile and a Gaussian of instrumental width (Gray 1992). Figure \ref{fig:lsd}b shows the result of this fitting procedure (in red smooth lines), as well as the synthetic profiles of both components. We then subtracted from the observed binary profile the synthetic secondary profile. The resulting observed primary profile is shown in Fig. \ref{fig:lsd}c.

\subsection{The oblique rotator model}

In order to characterise the magnetic fields of the HAeBe stars, we used the oblique rotator model described by Stift (1975). We consider a dipole placed at a distance $d_{\rm dip}$ on the magnetic axis of a spherical rotating star with a magnetic intensity at the pole $B_{\rm P}$. The rotation axis of the star is inclined at an angle $i$ with respect to the line of sight and makes an angle $\beta$ with the magnetic axis. 

In the weak field approximation, the Stokes $V$ profile amplitude is proportional to the magnetic field projected onto the line of sight and integrated over the surface of the star ($B_{\ell}$, the longitudinal magnetic field, hereafter). As the star rotates, the visible magnetic field changes, resulting in variation of $B_{\ell}$. Therefore the Stokes $V$ profile changes with the rotation phase.

\subsection{Stellar monitoring and fitting procedure}

In order to determine the geometrical and magnetic parameters $i$, $\beta$, $B_{\rm P}$ and $d_{\rm dip}$, as well as the rotation period $P$ of each star, we observed the stars at many different rotation phases and then fit simultaneously all the Stokes $V$ profiles observed for each star. With this aim we calculated a grid of $V$ profiles, using the oblique rotator model, for each date of observations, varying the five parameters. Then, for each star, we applied a $\chi^2$ minimisation to find the best model which matches simultaneously all the $V$ profiles observed. Fig. \ref{fig:fitv} shows the result of our fitting procedure for HD 200775. The synthetic Stokes $V$ are superimposed on the observed ones (Alecian et al. 2008a).

\section{Results and Discussion}

\begin{table}[t]
\caption{Fundamental, geometrical and magnetic parameters of the magnetic Herbig Ae/Be stars. References : 1 : \citet{alecian08a}, 2 : \citet{folsom08}, in press, 3 : \cite{catala07}.}
\label{tab:fp}
\centering
\begin{minipage}[t]{\linewidth}
\begin{tabular}{@{}lcccccccc@{}}
\hline
Star   & Sp. T & P    & $i$ ($^{\circ}$) & $\beta$ ($^{\circ}$) & $B_{\rm P}$ & $d_{\rm dip}$ & $B_{\rm P (ZAMS)}$ & Ref. \\
         &           & (d)  &                         &                                &  (kG)             & $R_*$             & (kG)                          &        \\
\hline
HD 200775 & B2 & 4.3281       & 60            & 125            & 1            & 0.05 & 3.6                                             & 1 \\
HD 72106   & A0 & 0.63995     & 23            & 60              & 1.3         & 0      & 1.3                                             & 2 \\
HD 190073%\footnote{Although we have observed HD 90073 over more that 2 years, no variation of the Stokes $V$ profile has yet been detected.} 
                   & A2 &                   & $[0,90]$   & $[0,90]$     &   $>0.3$ &         & $>1.2$                                       & 3 \\
\hline
\end{tabular}
\end{minipage}
\end{table}

The values of the geometrical and magnetic parameters are summarized in Table \ref{tab:fp} for the 3 stars studied to date. In the case of HD 190073, the topology and the intensity of its magnetic field are not well constrained. This is  because, during 3 years of observations, the Stokes $V$ profile has not been observed to vary. There are 3 possible explanations for this : the inclination $i$ is very small, the obliquity angle $\beta$ is very small, or the rotation period of the star is very long. More observations will allow us to discard two of these solutions. However the stability of the magnetic field over more than 2 years and the shape of the Stokes $V$ profiles lead us to the conclusion that this star hosts a large-scale fossil magnetic field (Catala et al. 2006).

The success of our fitting procedure for the stars HD 200775 and HD 72106, as well as our discussion on HD 190073, lead to the conclusion that the magnetic Herbig Ae/Be stars host globally dipolar magnetic fields, similar to the Ap/Bp stars.

Assuming the conservation of magnetic flux during PMS evolution, and using the current radii of the stars and their predicted radii on the ZAMS, we can estimate the magnetic intensity at their surface they will have when they will reach the ZAMS (see Table 1). We found intensities ranging from 300 G to 3.6 kG, which is very close to what is observed in the Ap/Bp stars. Hence we bring new strong arguments in favour of the fossil field hypothesis.
%\vspace{5mm}

%\section{Conclusion}

\acknowledgements %%% Text of acknowledgements runs on after this command.
EA is supported by the Marie Curie FP6 program.

%%% THE BIBLIOGRAPHY
%%%
%%% CONSULT SECTION 3 OF "INSTRUCTIONS FOR AUTHORS" FOR HOW TO USE NATBIB.
%%% AUTHORS ARE ENCOURAGED TO USE EITHER THE "THEBIBLIOGRAPY" ENVIRONMENT
%%% BY UNCOMMENTING (DELETING THE "%" SYMBOL) THE COMMANDS BELOW, OR BY
%%% USING THE BIBTEX ENVIRONMENT. TO FIND OUT WHICH IS APPLICABLE TO YOUR
%%% CONTRIBUTION, CONSULT THE VOLUME EDITORS FOR YOUR PROCEEDINGS.
%%%


\begin{thebibliography}{}
\bibitem[Alecian et al.(2008a)]{alecian08a}
     {{Alecian}, E., {Catala}, C., {Wade}, G.~A., et al.
     %{Donati}, J.-F., {Petit}, P., {Landstreet}, J.~D., {B\"ohm}, T., {Bouret}, J.-C., {Bagnulo}, S., {Folsom}, C., {Silvester}, J.
     }, 2008a, 
     \textit{MNRAS}, 385, 391

\bibitem[Alecian et al.(2008b)]{alecian2008b}
     {{Alecian}, E., {Wade}, G.~A., {Catala}, C., et al. 
     %{Bagnulo}, S., {Bohlender}, D., {B\"ohm}, T., {Bouret}, J.-C., {Donati}, J.-F., {Folsom}, C.P., {Grunhut}, J., {Landstreet}, J.~D.
     }, 2008b, 
      \textit{A\&A}, 481, L99 
      
\bibitem[Folsom et al. (2008)]{folsom08}
     {{Folsom}, C.P., {Wade}, G.A., {Kochukhov}, O., et al. 
     %{Alecian}, E., {Catala}, C., {Bagnulo}, S., {B\"om}, T., {Bouret}, J.-C., {Donati}, J.-F., {Grunhut}, J., {Landstreet}, J.D., {Hanes}, D.A.
     }, 2008, in press

\bibitem[Catala et al. (2007)]{catala07}
     {{Catala}, C., {Alecian}, E., {Donati}, J.-F., et al.
     %{Wade}, G.A., {Landstreet}, J.D., {B\"om}, T., {Bouret}, J.-C., {Bagnulo}, S., {Folsom}, C., {Silvester}, J.
     }, 2007,
     \textit{A\&A}, 462, 293 

\bibitem[Donati et al. (1997)]{donati97}
     {{Donati}, J.-F., {Semel}, M., {Carter}, B.~D., et al.
     %{Rees}, D.~E., {Collier Cameron}, A.
     }, 1997,
     \textit{MNRAS}, 291, 658

\bibitem[Wade et al. (2005)]{wade05}
     {{Wade}, G.~A., {Drouin}, D., {Bagnulo}, S., et al.
     %{Landstreet}, J.~D., {Mason}, E., {Silvester}, J., {Alecian}, E., {B{\"o}hm}, T., {Bouret}, J.-C., {Catala}, C., {Donati}, J.-F.
     }, 2005,
     \textit{A\&A}, 442, L31

\bibitem[Gray (1992)]{gray92}
     {Gray D.F.},
     {The Observation and Analysis of Stellar Photospheres},
     2nd edn. Cambridge

\bibitem[Kurucz (1993)]{kurucz93}
     {Kurucz R.}, 1993, Opacities for Stellar Atmospheres
     %: [?3.5], [?4.0], [?4.5]. Kurucz CD-ROM No. 7. Smithsonian Astrophysical Observatory, 
     Cambridge, MA, p. 7

\bibitem[Stift (1975)]{stift75}
     {{Stift}, M.~J.}, 1975,
     \textit{MNRAS}, 172, 133



\end{thebibliography}
\end{document}